\documentclass[amsmath,amsfonts,aps,nofootinbib,preprintnumbers]{revtex4}
\usepackage{graphicx}

\newcommand{\xpt}[0]{{$\chi$PT}}

\newcommand{\pqxpt}[0]{{PQ$\chi$PT}}

\begin{document}

\preprint{NT@UW-06-27}

\title{Generalised parton distributions of the pion in
  partially-quenched chiral perturbation theory}

\author{Jiunn-Wei Chen} \affiliation{Department of Physics and Center
  for Theoretical Sciences, National Taiwan University, Taipei,
  Taiwan}

\author{William Detmold} \affiliation{Department of Physics,
  University of Washington, Box 351560, Seattle, WA 98195, USA}

\author{Brian Smigielski} \affiliation{Department of Physics,
  University of Washington, Box 351560, Seattle, WA 98195, USA}

\begin{abstract}
  We consider the pion matrix elements of the isoscalar and isovector
  combinations of the vector and tensor twist-two operators that
  determine the moments of the various pion generalised parton
  distributions. Our analysis is performed using partially-quenched
  chiral perturbation theory. We work in the SU(2) and SU(4$|$2)
  theories and present our results at infinite volume and also at
  finite volume where some subtleties arise.  These results are useful
  for extrapolations of lattice calculations of these matrix elements
  at small momentum transfer to the physical regime.
\end{abstract}

\date\today \maketitle

\section{Introduction}
\label{sec:introduction}

Generalised parton distributions (GPDs)
\cite{Mueller:1998fv,Ji:1996ek,Ji:1996nm,Radyushkin:1997ki} provide a
uniquely detailed view of the structure of hadrons, unifying the
information encoded in form-factors and parton distributions, and
supplementing both. Ongoing experiments at DESY
\cite{Airapetian:2001yk,Adloff:2001cn,Chekanov:2003ya} and Jefferson
Lab \cite{Stepanyan:2001sm} (see
Refs.~\cite{Diehl:2003ny,Belitsky:2005qn} for recent reviews) seek to
learn about these fundamental quantities in deeply-virtual Compton
scattering (DVCS) and related processes.  Aspects of GPDs are also
being investigated in QCD and phenomenological models. Since GPDs
encode long distance hadronic structure, QCD analyses of them are
necessarily based on non-perturbative methods such as lattice QCD.
These examinations are complementary to the experimental efforts in
that different facets of the GPDs can be accessed. Most experimental
efforts are focused on the proton and either measure the GPDs at
constrained kinematics (through polarisation observables) or measure
integrals of the GPDs (through DVCS cross-sections).  Lattice QCD
analyses are based on the operator product expansion and lead to
information on the lowest few Mellin moments of the GPDs which
correspond to non-forward matrix elements of twist-two operators.
Again most studies focus on the proton
\cite{Hagler:2003jd,Gockeler:2003jf,Hagler:2003is,Gockeler:2005cj},
but the GPDs of other hadrons are equally accessible in the lattice
approach (unlike in experiment).  In particular, recent studies by the
QCDSF collaboration \cite{Brommel:2005ee} have investigated the GPDs
of the pion.

Here we investigate the vector and tensor GPDs of the pions from the
point of view of chiral perturbation theory applied to lattice QCD,
studying the quark mass and lattice volume dependence of the meson
matrix elements of twist-two operators.  Related studies of the vector
GPD pion matrix elements have been presented in
Ref.~\cite{Arndt:2001ye,Chen:2001eg,Kivel:2002ia,Detmold:2003tm,
  Chen:2003fp,Diehl:2005rn}, however we extend those results to
partially-quenched chiral perturbation theory (appropriate for lattice
calculations with differing sea- and valence- quark masses). We also
consider the tensor twist-two operators (also treated recently in
Ref.~\cite{Diehl:2006js}) and include the effects of the finite volume
to which lattice simulations are necessarily restricted (the Lorentz
non-invariance of the lattice boundary conditions introduces some
novel issues in the resulting extrapolation that we highlight).  The
finite volume calculations in partially-quenched chiral perturbation
theory are particularly relevant to the ongoing lattice calculations
reported in Ref.~\cite{Brommel:2005ee} (see \cite{Best:1997qp} for
earlier work in the forward limit).

In the following section we provide our notation and conventions for
the pion GPDs and twist-two matrix elements. In
Section~\ref{sec:effect-field-theory} we turn to the effective field
theory description of these objects before presenting our infinite
volume results in Section~\ref{sec:results}, their finite volume
analogues in Section~\ref{sec:finite-volume-funct}, and a concluding
discussion in Section~\ref{sec:discussion}. Various aspects of the
finite volume forms are relegated to the Appendix.

\section{Generalised parton distribution of the pion}
\label{sec:definitions}

\subsection{Pion GPDs}
\label{sec:pseudo-scalar-meson}

The generalised parton distributions of the pions (here we restrict
our discussion to SU(2)) are defined by matrix elements of light-cone
separated bi-local currents. Specifically, the vector GPDs $H^A$, and
the tensor GPDs, $E_{T}^A$, are given by
\begin{eqnarray}
\label{eq:1}
\langle \pi ^{i}(p^{\prime })|\,\overline{\psi}\left( -\frac{z}{2}
n\right) \tau ^{A } \gamma \cdot u \psi\left( \frac{z}{2}n\right) |\pi
^{j}(p)\rangle 
&=&\int_{-1}^{1} dye^{-iyzu\cdot \bar{P}}H^{A }(y,\xi,t)u\cdot
\bar{P}\text{tr}\left[ \ \!\!\tau ^{i}\tau ^{A }\tau 
^{j}\right] \ ,
\end{eqnarray}
and
\begin{eqnarray}
\label{eq:2}
\langle \pi ^{i}(p^{\prime })|\,\overline{\psi}\left( -\frac{z}{2}
n\right) \tau ^{A } i\, \sigma^{\mu\nu} u_\mu r_\nu \psi\left(
\frac{z}{2}n\right) |\pi 
^{j}(p)\rangle 
&=&\int_{-1}^{1} dy e^{-iyzu\cdot \bar{P}}E_{T}^{A }(y,\xi,t)u\cdot
\bar{P} r\cdot\Delta \text{tr}\left[ \ \!\!\tau ^{i}\tau ^{A
  }\tau 
^{j}\right] \ ,
\end{eqnarray}
respectively (for simplicity, we have suppressed the gauge links that
render these matrix elements gauge invariant). Here the average four
momentum of the incoming and outgoing pion states is
$\overline{P}=\frac{1}{2}(p+p^\prime)$ and the momentum transfer is
$\Delta=p^\prime-p$, $u$ is a light-like vector ($u^2=0$) and $r$ is
transverse ($r\cdot u=r\cdot\overline{P}=0$). The GPDs are functions
of the three variables $y$,
$\xi=-\frac{u\cdot\Delta}{2u\cdot\overline{P}}$ and $t=\Delta^2$.
Finally, $\tau_A=(\bf{1},\tau_{A})$ for QCD.

\subsection{Twist-two operators}
\label{sec:twist-two-operators}

As is evident from the forms of the GPDs, there are two towers of
local twist-two quark operators that have non-vanishing pion matrix
elements. These are given by
\begin{eqnarray}
  \label{eq:3}
  {\cal O}^{(n)}_A \equiv u_{\mu_0}\ldots u_{\mu_n} {\cal O}_A^{\mu_0\ldots
    \mu_n} = \overline\psi u\cdot\gamma(i\, u\cdot
  \tensor{D})^{n}\tau_A\psi\,,
\end{eqnarray}
and 
\begin{eqnarray}
  \label{eq:4}
   {\cal O}^{(n)}_{T;A} \equiv u_\alpha r_\beta u_{\mu_1}\ldots
   u_{\mu_{n}} {\cal 
     O}_{T;A}^{\mu_1\ldots\mu_n} = \overline\psi
   i\,\sigma^{\alpha\beta}u_\alpha r_\beta (i\, u\cdot
   \tensor{D})^{n}\tau_A\psi\,. 
\end{eqnarray}
The operators ${\cal O}_{A}^{\mu_1\ldots \mu_n}$ and ${\cal
  O}_{T;A}^{\mu_1\ldots \mu_n}$ are of fixed twist (= dimension -
spin) and transform irreducibly under the Lorentz group. Matrix
elements of the vector twist-two operators in Eq.~(\ref{eq:3}) give
moments of the quark distribution in the pion in the forward limit.
There are also two towers of purely gluonic operators at twist-two
(see for example, Ref.~\cite{Belitsky:2005qn}) that have non-zero
pionic matrix elements.  For the purposes of our current discussion we
note that the vector case has the same transformation properties as
${\cal O}^{(n)}_0$ above while the tensor case is beyond the scope of
this work.

These operators transform as $({\bf 3}, {\bf 1}) \oplus ({\bf 1}, {\bf
  3})$ (isovector, $A=1,2,3$) or $({\bf 1},{\bf 1})$ (isoscalar,
$A=0$) under SU(2)$_L\times$SU(2)$_R$ rotations. The tensor twist-two
operators also have non-zero matrix elements but vanish in the forward
limit and belong to the $({\bf 2}, \overline{\bf 2}) \oplus
(\overline{\bf 2}, {\bf 2})$ representation of
SU(2)$_L\times$SU(2)$_R$ irrespective of the flavour index $A$. In the
SU(4$|$2) partially-quenched QCD case (where additional valence and
ghost quarks are introduced), these operators are extended by the
replacement of $\tau_A$ by $\bar{\tau}_0$, $\bar\tau_a$ and
$\bar\tau_T$ in the isoscalar-vector, isovector-vector and tensor
cases. These matrices are somewhat arbitrary
\cite{Chen:2001yi,Beane:2002vq,Detmold:2005pt}, but for definiteness
we choose:
\begin{eqnarray}
  \label{eq:5}
  \bar\tau_0={\rm diag}(1,1,1,1,1,1)\,,
  \quad\quad
  \bar\tau_3={\rm diag}(1,-1,q_j,q_k,q_l,q_j+q_k-q_l)\,,
  \quad\quad
  \bar\tau_T={\rm diag}(1,y,q_3,q_4,q_5,q_3+q_4-q_5)\,,
\end{eqnarray}
which reduce to the usual Pauli matrices in the QCD limit and
transform in the corresponding representations of the enlarged flavour
group.

The local twist-two QCD operators are simply related to those in Eqs.
(\ref{eq:1}) and (\ref{eq:2}) and it follows that
\begin{eqnarray}
  \label{eq:6}
  \langle \pi ^{i}(p^{\prime })|\mathcal{O}_{A}^{(n)}|\pi
  ^{j}(p)\rangle 
  &=&H_{n+1}^{A }(\xi,t) \ (u\cdot \bar{P})^{n+1} \ \text{tr}\left[ \
    \!\!\bar\tau ^{i}\bar\tau ^{A }\bar\tau ^{j}\right]  \,,
  \\
  \langle \pi ^{i}(p^{\prime })|\mathcal{O}_{T,A}^{(n)}|\pi
  ^{j}(p)\rangle 
  &=&E_{T,n+1}^{A }(\xi,t) \ (r \cdot \Delta) \  (u\cdot \bar{P})^{n+1}
  \ \text{tr}\left[ \ \!\!\bar\tau ^{i}\bar\tau ^{A }\bar\tau
    ^{j}\right] \,, 
\end{eqnarray}
where $H_{n+1}^{A }(\xi,t) = \int_{-1}^{1} dy y^{n} H^{A }(y,\xi,t)$
and $E_{T,n+1}^{A}(\xi,t) = \int_{-1}^{1} dy y^{n}
E_{T}^{A}(y,\xi,t)$.

Discrete symmetries and the approximate isospin symmetry of QCD
constrain the pion GPDs. Time reversal invariance demands $H^{A
}(y,\xi,t)=H^{A }(y,-\xi,t)$ and $E_{T}^{A
}(y,\xi,t)=E_{T}^{A}(y,-\xi,t)$. Under charge conjugation
($\mathcal{C}$), both the vector and tensor operator transform as
\cite{Chen:2003fp,Ando:2006sk}:
\begin{eqnarray}
\mathcal{C}\mathcal{O}_{A}^{(n)}\mathcal{C}^{-1} &=& 
(-1)^{n+1}\mathcal{O}_{A}^{(n)}\,,
\\
\mathcal{C}\mathcal{O}_{T;A}^{(n)}\mathcal{C}^{-1} &=& 
(-1)^{n+1}\mathcal{O}_{T;A}^{(n)}\,.
\end{eqnarray}
Using this, it can be shown that the isoscalar ($A=0$) vector matrix
elements vanish for even index $n=2k$ and the isovector ($A=3$) vector
matrix elements vanish for odd $n$ (additional complications arise in
the SU(4$|$2) case). For odd index, the isoscalar-vector matrix
elements are parameterised in terms of generalised form factors
$A_{n,j}^{(0)}$ and $C_{n+1}^{(0)}$ as:
\begin{eqnarray}
\label{eq:7}
\langle \pi ^{i}(P^{\prime })|\mathcal{O}_{0}^{(2k-1)}|\pi
^{j}(P)\rangle
&=&2\delta _{ij}\left\{ \sum_{l=0}^{k-1}\left( u\cdot \Delta \right)
^{2l}\left( u\cdot \bar{P}\right) ^{2k-2l}A_{2k,2l}^{(0)}(t)+\left(
  u\cdot \Delta \right) ^{2k}C_{2k}^{(0)}(t)\right\} \, , 
\end{eqnarray}
while for even index, the isovector-vector matrix element can be
parameterised as: 
\begin{eqnarray}
\label{eq:8}
\langle \pi ^{i}(P^{\prime })|O_{j}^{(2k)}|\pi ^{k}(P)\rangle
&=&2i\epsilon ^{ijk}\sum_{l=0}^{k}\left( u\cdot \Delta \right)
^{2l}\left( u\cdot \bar{P}\right) ^{2k-2l+1}A_{2k+1,2l}^{(3)}(t)\, .
\end{eqnarray}

Similarly, parameterisations of the tensor operator matrix elements
are given by
\begin{eqnarray}
  \label{eq:9}
\langle \pi ^{i}(p^{\prime })|{\cal O}^{(2k+1)}_{T;0}|\pi ^{j}(p)\rangle 
&=&2\delta _{ij}\sum_{l=0}^{k}\frac{1}{\Lambda_\chi} \left(
    u\cdot \overline{P} 
\right) \left( r\cdot \Delta \right)
\left( u\cdot \Delta
\right) ^{2l}\left( u\cdot \overline{P}\right)
^{2k-2l+1}B_{2k+2,2l}^{T,(0)}(t)\,,
 \\
  \label{eq:10}
\langle \pi ^{i}(p^{\prime })|{\cal O}^{(2k)}_{T;3}|\pi ^{j}(p)\rangle
&=&2i\epsilon ^{i3j}\sum_{l=1}^{k}\frac{1}{\Lambda_\chi} \left(
    u\cdot \overline{P}\right) \left( r\cdot \Delta \right)
\left( u\cdot \Delta
\right) ^{2l}\left( u\cdot \overline{P}\right)
^{2k-2l}B_{2k+1,2l}^{T,(3)}(t)\,,
\end{eqnarray}
for both even and odd index.

The generalised form factors, $A^{(A)}_{n,k}$ and $B^{T,(A)}_{n,k}$,
are related to the generalised parton distributions in Eq.~(\ref{eq:1})
and (\ref{eq:2}) as
\begin{eqnarray}
  \label{eq:11}
 \int_{-1}^{1} dy \ y^{n} H^{(0)}(y,\xi,t)&=&\sum_{j=0,
   \text{even}}^{n-1}(-2 \xi)^{j}A^{(0)}_{n+1,j}(t)+(-2
 \xi)^{n+1}C_{n+1}^0(t) \,,
 \\
  \int_{-1}^{1} dy \ y^{n} H^{(3)}(y,\xi,t)&=&\sum_{j=0,
    \text{even}}^{n}(-2 \xi)^{j}A^{(3)}_{n+1,j}(t) \,,
  \end{eqnarray}
and 
\begin{eqnarray}
  \label{eq:12}
 \int_{-1}^{1} dy \ y^{n} E_{T,n+1}^{A}(y, \xi,
 t)=\sum_{j=0,\text{even}}^{n}(2 \xi)^{j}B^{T,(A)}_{n+1,j}(t)\,, 
\end{eqnarray}
respectively.  $C^{(0)}_{n+1}(t)$ is the D-term form-factor
\cite{Polyakov:1999gs}. 

\section{Effective field theory}
\label{sec:effect-field-theory}

The small $t$ behaviour of the GPD form-factors can be reliably
described by the low energy effective theory of QCD, chiral
perturbation theory (\xpt) \cite{Weinberg:1978kz,Gasser:1983yg}. The
extensions of this to partially-quenched QCD\footnote{We do not
  discuss quenched QCD in which sea quarks are omitted as it has no
  connection to physical observables except in the large $N_c$ limit
  \cite{Chen:2002mj}.} (in which valence and sea quarks have different
masses as appropriate for many current lattice calculations),
partially-quenched \xpt\ (PQ\xpt) \cite{Bernard:1993sv,Sharpe:1997by},
is well known and here we simply highlight the relevant pieces of the
Lagrangian and discuss the operators that contribute to the twist-two
matrix elements.  We primarily focus on a partially-quenched theory of
valence ($u$, $d$), sea ($j$, $l$) and ghost ($\tilde u,\,\tilde d$)
quarks with masses contained in the matrix
\begin{equation}
\label{eq:Mq_def}
m_Q = {\rm  diag}(m_u,m_d,m_j,m_l,m_{\tilde u},m_{\tilde d})\,,
\end{equation}
where $m_{\tilde u,\tilde d}=m_{u,d}$ such that the path-integral
determinants arising from the valence and ghost quark sectors exactly
cancel.

\subsection{Lagrangian}
\label{sec:lagrangian}

At leading order the \pqxpt\ Lagrangian is given by
\begin{eqnarray}
{\cal L }_\Phi & = & 
{f^2\over 8} 
{\rm str}\left[ {\cal D}^\mu\Sigma^\dagger {\cal D}_\mu\Sigma \right] + 
\lambda {f^2\over 4} 
{\rm str}\left[ m_Q\Sigma^\dagger + m_Q^\dagger\Sigma \right],
\label{eq:PGBlagrangian}
\end{eqnarray}
where the pseudo-Goldstone mesons are embedded non-linearly in the
coset field
\begin{equation}
\label{eq:Sigma_def}
\Sigma = \exp{\left(\frac{2\,i\,\Phi}{f}\right)} ,
\end{equation}
(under a chiral rotation, $\Sigma\to L\Sigma R^\dagger$) with the
matrix $\Phi$ given by
\begin{equation}
  \label{eq:Phi_def}
\Phi = \begin{pmatrix} M & \chi^\dagger \cr \chi &\tilde{M} \end{pmatrix}
\,,
\end{equation}
and
\begin{eqnarray}
  \label{eq:Mchi_def}
        M =\begin{pmatrix}
                \eta_u & \pi^+ & \phi_{uj} & \phi_{ul} \\
                \pi^- & \eta_d & \phi_{dj} & \phi_{dl} \\
                \phi_{ju} & \phi_{jd} & \eta_j & \phi_{jl} \\
                \phi_{lu} & \phi_{ld} & \phi_{lj} & \eta_l
                \end{pmatrix}\ ,
&  \hspace*{15mm}
\tilde{M}=\begin{pmatrix} \tilde\eta_u    & \tilde\pi^+        \\
                          \tilde\pi^-     & \tilde\eta_d       
                          \end{pmatrix}\,, 
\hspace*{15mm} &
        \chi=\begin{pmatrix} 
                \phi_{\tilde{u}u} & \phi_{\tilde{u}d} & \phi_{\tilde{u}j} & \phi_{\tilde{u}l} \\
                \phi_{\tilde{d}u} & \phi_{\tilde{d}d} & \phi_{\tilde{d}j} & \phi_{\tilde{d}l} 
                \end{pmatrix}\,.
\end{eqnarray}
The upper left $2\times2$ block of $M$ corresponds to the usual
valence--valence mesons, the lower right to sea--sea mesons and the
remaining entries of $M$ to valence--sea mesons. Mesons in $\tilde{M}$
are composed of ghost quarks and ghost anti-quarks and are thus
bosonic.  Mesons in $\chi$ contain ghost--valence or ghost--sea
quark--anti-quark pairs and are fermionic. In terms of the quark
masses, the tree-level meson masses are given by
\begin{eqnarray}
  \label{eq:mesonmass_def}
  M^2_{\Phi_{ij}}=M_{{ Q}_i { Q}_j}^2 = 
\lambda\left[ \left(m_Q\right)_{ii}
    + \left(m_Q\right)_{jj} \right] \,,
\end{eqnarray}
where ${ Q}=(u,\,d,\,j,\,l,\,\tilde{u},\,\tilde{d})$. The decay
constant is normalised as $f\sim132$~MeV.
Additional terms involving the flavour singlet field, str[$\Phi$] are
not relevant here; in both \pqxpt\ and \xpt\, the singlet meson
acquires a large mass through the strong U(1)$_A$ anomaly and can be
integrated out, leading to a modified flavour neutral propagator that
contains both single and double pole
structures~\cite{Sharpe:2000bc,Sharpe:2001fh}.

\subsection{Twist-two operators}

Twist-two operators have been studied quite extensively in various low
energy effective theories.  A number of studies have focused on pionic
matrix elements of twist-two operators \cite{Arndt:2001ye,Chen:2001eg,
  Kivel:2002ia,Detmold:2003tm,Chen:2003fp,Diehl:2005rn} but the
relevant operators also contribute in numerous studies of nucleon
matrix elements \cite{Detmold:2001jb,Chen:2001pv,
  Belitsky:2002jp,Detmold:2002nf,Detmold:2005pt,Ando:2006sk,
  Diehl:2006ya,Diehl:2006js} (these studies have also been extended to
the nuclear setting in Refs.~\cite{Beane:2004xf,Chen:2004zx}).

We first focus on the vector operators, Eq.~(\ref{eq:3}). To perform
the matching of these operators to those in $\chi$PT it is useful to
make the separation:
\begin{eqnarray}
\mathcal{O}_{A}^{(m)}=\mathcal{O}_{A,L}^{(m)}+\mathcal{O}_{A,R}^{(m)}\,,
\end{eqnarray}
such that $\mathcal{O}_{A,H}^{(m)}=\bar{q}_{H}\bar\tau^{A}_{H}(u \cdot
\gamma)(u \cdot i\tensor{D})^{m-1}q_{H}$ for $H=L,R$, where
$q_{L,R}=[(1 \mp \gamma_{5})/2]q$ are the left (right)-handed quark
fields which transform as $q_{L} \rightarrow Lq_{L}$ and $q_{R}
\rightarrow Rq_{R}$ under the action of SU(4$|$2)$_{L}\times
$SU(4$|$2)$_{R}$. To construct the EFT operators, it is useful to
treat $\bar\tau_{L,R}^{A}$ as a spurion field that transform under
global chiral rotations as:
\begin{eqnarray}
\bar\tau_{L}^{A} \rightarrow L\bar\tau_{L}^{A}L^{\dagger}, 
\ \  \bar\tau_{R}^{A} \rightarrow R\bar\tau_{R}^{A}R^{\dagger}
\end{eqnarray}
(the spurion fields take a vacuum expectation value of $\bar\tau_A$).
This promotion renders the QCD operators, $\mathcal{O}_{A}^{(m)}$,
invariant under chiral rotations.  For these operators to have the
correct charge conjugation properties, Eq.(\ref{eq:9}), requires
$\bar\tau_{L,R}^{A} \stackrel{\cal C}{\rightarrow}
(\bar\tau_{L,R}^{A})^{T}$ since $\Sigma \stackrel{\cal C}{\rightarrow}
\Sigma^{T}$ and $m_{q} \stackrel{\cal C}{\rightarrow} (m_{q})^{T}$.

At leading order, the EFT operators consistent with these
transformation properties are constructed from a single insertion of
$\tau_{L,R}^A$ and arbitrary numbers of $\Sigma$/$\Sigma^{\dagger}$
pairs. The first two terms in this tower are given by:

\begin{eqnarray}
&&\mathcal{O}_{A,1}^{(m)}=\sum_{k=0}^{m+1} \left(a_{m,k}^{A,1} \
  \text{str}[\bar\tau_{A}(\square^{k}\Sigma)(\square^{m-k+1}
  \Sigma^{\dagger})]+\ a_{m,k}^{A,1a} \
  \text{str}[\bar\tau_{A}(\square^{k}
  \Sigma^{\dagger})(\square^{m-k+1}\Sigma)] \right)   \,,
\label{eq:13}
\\
\label{eq:14}
&&\mathcal{O}_{A,2}^{(m)}=\sum_{j_{1},j_{2},j_{3}=0}^{j_{1}+j_{2}+j_{3}<m+1}
a_{m,j_{1},j_{2},j_{3}}^{A,2} \
\text{str}[\bar\tau_{A}(\square^{j_{1}}\Sigma)
(\square^{j_{2}}\Sigma^{\dagger})(\square^{j_{3}}\Sigma) 
(\square^{m+1-j_{1}-j_{2}-j_{3}}
\Sigma^{\dagger})+(\Sigma \leftrightarrow \Sigma^{\dagger})] \,,
\end{eqnarray}
where $\square^{k}=(i u \cdot \partial)^{k}$.  Charge conjugation
requires that $a_{m,m-k+1}^{A,1a}=a_{m,k}^{A,1}(-1)^{m+1}$, and
parity, $\Sigma \rightarrow \Sigma^{\dagger}$, implies that
$a_{m,k}^{A,1} = (-1)^{m+1} a_{m,m-k+1}^{A,1}$, significantly
simplifying $\mathcal{O}_{A,1}^{(m)}$. The second type of operator,
$\mathcal{O}_{A,2}^{(m)}$, is not independent of
$\mathcal{O}_{A,1}^{(m)}$ at infinite volume. If there are non-zero
numbers of derivatives on three or four coset fields
($\Sigma^{(\dagger)}$), this operator may contribute through diagram
(b) in Figure \ref{fig:XPT-diagrams} below, but will be proportional
to powers of $u \cdot k$ where $k$ is the integration momentum. For
even powers this will produce overall factors of $u^{2}$ which vanish;
for odd powers, the integrand will be odd and hence vanish upon
integration. Consequently, these operators only contribute in
Fig.\ref{fig:XPT-diagrams}(b) when all derivatives act on the external
pion fields. However using integration by parts and eliminating
operators with derivatives on more than two meson fields such terms
can be rewritten in terms of $\mathcal{O}_{A,1}^{(m)}$.  Operators
involving six or more coset fields can similarly be eliminated.

Thus the final form for the vector twist-two operators we use is:
\begin{eqnarray}
  \label{eq:15}
  {\cal O}_{A}^{(m)} &=&
\frac{f^2}{4}\sum_{j=0,\,{\rm even}}^{m} a_{m+1,j}^{A}
(-\square)^j\Big\{ 
{\rm str}\left[ \bar\tau_A \Sigma\,
    \tensor\square^{m+1-j} \Sigma^\dagger\right]+
{\rm str}\left[
  \bar\tau_A\, \Sigma^\dagger\tensor\square^{m-j+1}\Sigma\right] 
\Big\}\,,
\end{eqnarray}
where we find it convenient to express the result in terms of the
forward-backward derivative $\tensor\square = \roarrow\square
-\loarrow\square = i\,u\cdot(\roarrow{\partial} - \loarrow{\partial})$
and time-reversal invariance limits the sum to even
values.\footnote{For notational convenience, we use the same symbol to
  denote both the underlying QCD operator and that in the effective
  theory. Note, however, the EFT operators match not only the leading
  twist QCD operators but also higher twist operators of the same
  quantum numbers and also to purely gluonic operators in the
  isoscalar cases \cite{Chen:2004zx}.}  At finite volume some of the
operators we have neglected will contribute as Lorentz symmetry is no
longer preserved.  Here we ignore such terms but they result in
additional complications in the extrapolation needed for lattice data
as discussed in Section \ref{sec:finite-volume-funct}.

Construction of the tensor operators is similar to that of the vector
operators, however in analysing the transformation properties of the
QCD operators it is necessary to introduce additional spurion fields,
$\bar\tau_{LR}^{A}$ and $\bar\tau_{RL}^{A}$, that transform as
$\bar\tau_{LR}^{A} \rightarrow L\bar\tau_{LR}^{A}R^{\dagger}$, and
$\bar\tau_{RL}^{A} \rightarrow R\bar\tau_{LR}^{A}L^{\dagger}$ such
that
\begin{eqnarray}
\mathcal{O}_{T,A}^{(m)}=\bar{q}_{L}\bar\tau_{LR}^{A}i
u_{\mu}r_{\nu}\sigma^{\mu\nu}(iu \cdot
D)^{m}q_{R}+\bar{q}_{R}\bar\tau_{RL}^{A}i u_{\mu}r_{\nu}\sigma^{\mu\nu}(iu
\cdot D)^{m}q_{L}\,,
\end{eqnarray}
is invariant under chiral rotations.

As in the vector case, there is a tower of operators at leading order
in the EFT with arbitrary odd numbers of $\Sigma$ and
$\Sigma^{\dagger}$ fields and a single insertion of
$\bar\tau_{LR,RL}^{A}$consistent with the requisite transformation
properties.  A similar discussion to that above for the vector
operators greatly simplifies the structure of these operators and
shows that only the operators involving three coset fields contribute
to the single-particle matrix elements we are considering at
next-to-leading order in infinite volume.  Using charge conjugation
and noting the QCD operator is antisymmetric under the interchange $u
\leftrightarrow r$, this relevant operator can be written as:
\begin{eqnarray}
  \label{eq:16}
  {\cal O}_{T,A}^{(m)} &=& 
\frac{f^2}{2}\sum_{j=0,\,{\rm even}}^{m} b_{m+1,j} (- \square)^j\Big\{ 
(-\bar\square)\, {\rm str}\left[ \bar\tau_T \left\{\Sigma
    \left(\Sigma^\dagger 
    \,\tensor\square^{m+1-j} \Sigma \right)+\left(\Sigma
    \,\tensor\square^{m+1-j} \Sigma^\dagger \right)\Sigma
\right\}\right] 
\\
&&\hspace*{1cm}
- (-\square)\,{\rm str}\left[\bar \tau_T
  \left\{\Sigma
    \left(\Sigma^\dagger \,\tensor{\bar\square}\,
      \tensor\square^{m-j} \Sigma\right) +
    \left(\Sigma  \,\tensor{\bar\square}\,
      \tensor\square^{m-j}  \Sigma^\dagger\right) \Sigma \right\}\right]
+(\Sigma\leftrightarrow\Sigma^\dagger) 
\Big\}\,,
\nonumber
\end{eqnarray}
where $\bar\square=(i r \cdot \partial)$.  

Unlike the vector operators, the tensor operators involve a single set
of low energy constants (LECs) because they belong to the same chiral
representation regardless of the choice of flavour structure.  In the
SU(2) case, the super-traces in the operators of Eq.~(\ref{eq:15}) and
Eq.~(\ref{eq:16}) reduce to ordinary flavour traces and the various
matrices are now $2\times2$, but the form of the operators and the
LECs that appear are otherwise unchanged.

At the next-to-leading order (NLO), ${\cal O}(p^2)$, vector and tensor
operators which will contribute at tree-level are generated by
combining the leading order operators above with insertions of
$\partial^2$, $\partial^\alpha\ldots\partial_\alpha$ or by
substitution of the quark mass matrix $m_Q$ for a coset field,
$\Sigma$. The explicit forms of these operators are not required here
however they generate polynomial dependence on the quark masses and
$t$.

\section{Infinite volume results}
\label{sec:results}

\begin{figure}[!t]
  \centering
\begin{tabular}{ccccc}
  \includegraphics[width=0.25\columnwidth]{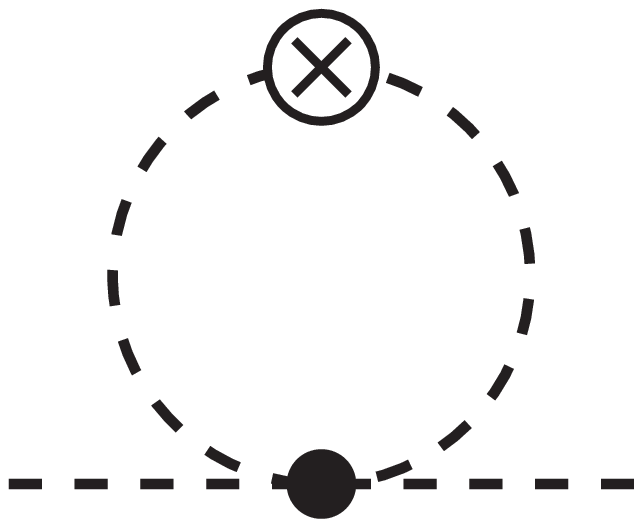}
& \hspace*{5mm}
&  \includegraphics[width=0.25\columnwidth]{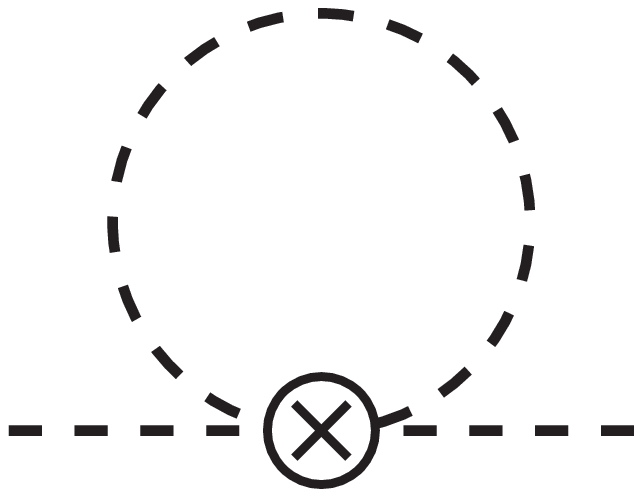}
& \hspace*{5mm}
&  \includegraphics[width=0.25\columnwidth]{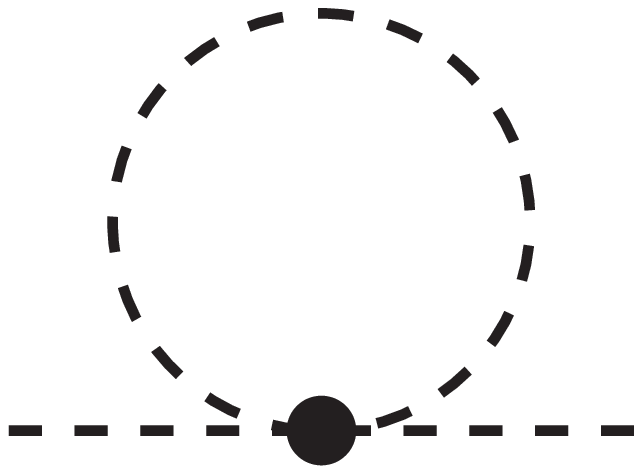} \\
(a) & \hspace*{5mm} & (b) & \hspace*{5mm} & (c) 
\end{tabular}
  \caption{One loop diagrams contributing to the mesonic matrix
    elements of twist-two operators. The filled circle denotes a
    vertex from the chiral Lagrangian while the crossed circle
    corresponds to the twist-two operator. Diagram (c) denotes the
    wave function renormalisation.} 
  \label{fig:XPT-diagrams}
\end{figure}

At leading order, the moments of the pion GPDs receive tree-level
contributions from the operators in Eqs.~(\ref{eq:15}) and
(\ref{eq:16}) above. At next-to-leading order contributions come from
tree-level insertions of the ${\cal O}(p^2)$ operators discussed above
and from the one-loop diagrams involving the leading order operators
shown in Fig.~\ref{fig:XPT-diagrams}. The higher order operators lead
to polynomial dependence of the GPD form-factors on $m_q\sim M^2$ (M
is a Goldstone meson mass) and $t$ whilst the loops generate
non-analytic dependence on these quantities.
Therefore at NLO, the vector GPD form factors will have the form:
\begin{eqnarray}
A^{(A)}_{m+1,j}(t)=A^{(A)}_{m+1,j}+A^{(A,a)}_{m+1,j}M^2+A^{(A,b)}_{m+1,j}t
+\text{loop\, contributions}, 
\end{eqnarray}
where $A^{(A)}_{m+1,j}$ is the bare matrix element determined in terms
of the leading order LECs $a_{m+1,j}^{A}$, and $A^{(A,a)}_{m+1,j}$ and
$A^{(A,b)}_{m+1,j}$ can be expressed in terms of linear combinations
of the LECs accompanying the various NLO operators and absorb
divergences from the loop contributions (in general this renders these
terms renormalisation scale dependent). A similar expression holds for
the tensor GPD form factors.

In the following subsections, we present results for the various
matrix elements in the SU(4$|$2) isospin limit case, $m_d=m_u$ and
$m_l=m_j$.  In this limit, only the valence-valence meson mass,
$M_{uu}=M_\pi$, and the valence-sea meson mass,
$M_{uj}^2=M_\pi^2-\frac{1}{2}\delta_{uj}^2$, enter. Here
$\delta_{uj}^2/2\lambda$ is the mass splitting between $u$ and $j$
quarks and results in SU(2) are easily obtained by setting
$\delta_{uj}\to0$ (these results are given in Appendix
\ref{sec:qcd-form-factors}).

\subsection{Isoscalar-vector operators}
\label{sec:isosc-vect-oper}

In the partially-quenched theory, the pion matrix elements of the
isoscalar-vector operators have the form
\begin{eqnarray}
  \label{eq:17}
\langle \pi^+|{\cal O}^{(m)}_0|\pi^+\rangle &=& 
2\left(1-(-1)^m\right)\sum_{j=0,\rm even}^{m}a^{(0)}_{m+1,j}
\left[(2\overline{P}\cdot u)^{m-j+1}(\Delta\cdot u)^j-(\Delta\cdot
  u)^{m+1}\right]
\\&&
-\frac{\left(1-(-1)^m\right)}{48 f^2 \pi ^2}
 (u\cdot \Delta )^{m+1}\sum_{j=0,\rm even}^{m}a^{(0)}_{m+1,j}
 \int_{-1}^{1}d\alpha
\Bigg\{\frac{12
      \left(1-\alpha ^{m+1-j}\right) \delta _{{uj}}^2 M_{\pi
      }^2}{4 
   M_{\pi }^2+t \left(\alpha ^2-1\right)}
\nonumber
\\&&\nonumber\hspace*{0.5cm}
+3 \left(1-\alpha
   ^{m+1-j}\right) \log \left[\frac{M_{\pi }^2+\frac{1}{4} t
     \left(\alpha 
   ^2-1\right)}{\mu ^2}\right] M_{\pi }^2
+\log
\left[\frac{M_{uj}^2+\frac{1}{4} t \left(\alpha ^2-1\right)}{\mu
    ^2}\right]
\\&&\hspace*{0.5cm}
\times \left(t 
   \left((j-m-4) \alpha ^2-j+m+8\right) \alpha ^{m+1-j}+4
   \left((j-m-2) \alpha ^{m+1-j}+1\right) M_{{uj}}^2+t \left(3
     \alpha 
   ^2-7\right)\right)
\nonumber
\Bigg\}  \,,
\end{eqnarray}
for $m$ odd and vanish for $m$ even. Note that above and in what
follows, we have suppressed analytic dependence of matrix elements on
$M_\pi^2$, $M_{uj}^2$ and $t$. The first term in the braces arises
from a double-pole propagator in Figure~\ref{fig:XPT-diagrams}(a) and
gives enhanced quark mass dependence. The QCD limit is easily obtained
by setting $\delta_{uj}=0$ and $M_{uj}\to M_\pi$.

Decomposing this result leads to the following structure for the GPD
form factors ($m$ odd):
\begin{eqnarray}
  \label{eq:18}
 A^{(0)}_{m+1,j}(t)&=&A^{(0)}_{m+1,j}+A^{(0,a)}_{m+1,j}M_\pi^2
+\tilde{A}^{(0,a)}_{m+1,j}M_{uj}^2
+A^{(0,b)}_{m+1,j}t, \ \text{for} \ j \leq m \,,
\end{eqnarray}
with no non-analytic dependence, and ($m$ odd)
\begin{eqnarray}
C^{(0)}_{m+1}(t)&=&C^{(0)}_{m+1}
+\frac{C^{(0)}_{m+1}}{96\pi^2f^2}\int_{-1}^{1} 
d\alpha \ \Bigg\{
\frac{12\delta_{uj}^2M_\pi^2}{4M_\pi^2+t(\alpha^2-1)} +3 M_\pi^2
\log\left[\frac{M_{\pi}^2-\frac{t}{4}(1-\alpha^2)}{\mu^2} 
\right]
\\&&\hspace*{1cm}
+((3\alpha^2-7)t+4M_{uj}^2)
\log\left[\frac{M_{uj}^2-\frac{t}{4}(1-\alpha^2)}{\mu^2}
\right]
\Bigg\}
+C^{(0,a)}_{m+1}M_\pi^2+\tilde{C}^{(0,a)}_{m+1}M_{uj}^2+C^{(0,b)}_{m+1}t 
\nonumber  \\ &&
+\frac{1}{192 \pi^2 f^2}\sum_{j=0,\text{even}}^{m}
2^{j-m} A^{(0)}_{m+1,j} \int_{-1}^{1} d\alpha \
\alpha^{m-j+1}
\Bigg\{
\frac{12\delta_{uj}^2M_\pi^2}{4M_\pi^2+t(\alpha^2-1)}
+3 M_\pi^2
\log\left[\frac{M_{\pi}^2-\frac{t}{4}(1-\alpha^2)}{\mu^2}\right]
\nonumber \\&&\hspace*{1cm}
+\left(t(3\alpha^2-7)+4M_{uj}^2+4(m-j+1)
\left(M_{uj}^2-\frac{t}{4}(1-\alpha^2)\right)\right)
\log\left[\frac{M_{uj}^2-\frac{t}{4}(1-\alpha^2)}{\mu^2}\right]
\Bigg\}\,,
\nonumber
\end{eqnarray}
where the LECs and the bare form-factors are related by
$C_{m+1}^{(0)}=-\sum_{j=0,\text{even}}^{m}a_{m+1,j}^{(0)}$ and
$A_{m+1,j}^{(0)}=2^{m-j+1}a_{m+1,j}^{(0)}$.  The $\pi^-$ and $\pi^0$
results are identical and, with appropriate changes in normalisation,
the QCD limits of these results agree with those in
Refs.~\cite{Kivel:2002ia,Diehl:2005rn} using integration by parts and
noting that $\sum_{j,{\rm even}} 2^j A^{(0)}_{m+1,j} = 0$ (the results for
the gravitational form-factors, $A_{2,0}(t)$ and $C_2(t)$, also agree
with previous calculations \cite{Donoghue:1991qv}). In the partially
quenched case, this form-factor leads to an enhanced divergence in the
pion gravitational radius $\sim \delta_{uj}^2/M_\pi^2$ as opposed to
$\ln M_\pi$ in QCD.

\subsection{Isovector-vector operators}
\label{sec:isov-vect-oper}

The pion matrix elements of the isovector-vector operators have the form
\begin{eqnarray}
  \label{eq:19}
 \langle \pi^+|{\cal O}^{(m)}_3|\pi^+\rangle &=&
 (1+(-1)^m)\sum_{j=0,\text{even}}^{m}a_{m+1,j}^{(3)}
 \Bigg\{2(2\overline{P} \cdot u)^{m-j+1} 
 (\Delta \cdot u)^j\left(1- 
  \frac{1}{8\pi^2 f^2}M_{uj}^2 \ \log\left[\frac{M_{uj}^2}{\mu^2}
  \right]\right)
 \\ \nonumber  \hspace*{1cm}
 &&  -\frac{1}{8\pi^2 f^2} \ (2\overline{P} \cdot u)(\Delta \cdot
 u)^{m}
\left(\frac{t}{2}\int_{-1}^{1} d\alpha \ \alpha^{m-j+2} \
 \text{log}\left[\frac{M_{uj}^2-\frac{t}{4}(1-\alpha^2)}{\mu^2} \right]
-2M_{uj}^2\text{log}\left[\frac{M_{uj}^2}{\mu^2}\right]\right)
\Bigg\} \,,
\end{eqnarray}
for even $m$. This leads to the following structure for the GPD form
factors:
\begin{eqnarray}
  \label{eq:20}
  A^{(3)}_{m+1,j}(t) &=& A^{(3)}_{m+1,j}
  \left(1-\frac{M_{uj}^2}{8\pi^2f^2}
    \text{log}\left(\frac{M_{uj}^2}{\mu^2}\right)
  \right)+A^{(3,a)}_{m+1,j}M_\pi^2 
+\tilde{A}^{(3,a)}_{m+1,j}M_{uj}^2
+A^{(3,b)}_{m+1,j}t, \ \text{for} \ j 
  \leq m-2  \,,
\end{eqnarray}
and
\begin{eqnarray}
 A^{(3)}_{m+1,m}(t) &=& A^{(3)}_{m+1,m}
 \left(1-\frac{M_{uj}^2}{8\pi^2f^2}
   \text{log}\left(\frac{M_{uj}^2}{\mu^2}\right)\right) 
+A^{(3,a)}_{m+1,m}M_\pi^2
+\tilde{A}^{(3,a)}_{m+1,m}M_{uj}^2
+A^{(3,b)}_{m+1,m}t
 \\&& \nonumber
 +\sum_{j=0,\text{even}}^{m} \frac{2^{j-m}A_{m+1,j}^{(3)}}{8\pi^2 f^2}
\left\{M_{uj}^2 \text{log} 
\left[\frac{M_{uj}^2}{\mu^2}\right]
  -\frac{t}{4}
  \int_{-1}^{1}
 d\alpha \ \alpha^{m-j+2} \
 \text{log}\left[\frac{M_{uj}^2-\frac{t}{4}(1-\alpha^2)}{\mu^2}
 \right]\right\}\,, 
\end{eqnarray}
using $a_{m+1,j}^{(3)}=2^{j-m-1}A_{m+1,j}^{(3)}$.  The $\pi^-$ results
are related to these by factors of $-1$ and those in the $\pi^0$
vanish.
The LECs $A_{1,0}^{(3,a)}$ and $\tilde{A}_{1,0}^{(3,a)}$ vanish by
current conservation and $A_{1,0}^{(3)}=1$.
These results can be shown to agree with
Ref.~\cite{Diehl:2005rn,Kivel:2002ia} (and earlier results in the case
of the vector-isovector form-factor, $A_{1,0}^{(3)}(t)$) using
integration by parts and noting the different normalisations.

In the partially-quenched theory, isospin is not a good quantum number
(the SU(4$|$2) adjoint matrices are given in Eq.~(\ref{eq:5})) and the
odd-$m$ matrix elements are also non-zero. These take the form
($\pi^-$ and $\pi^0$ are identical to the $\pi^+$)
\begin{eqnarray}
  \label{eq:21}
 \langle \pi^+|{\cal O}^{(m)}_3|\pi^+\rangle_{m\rm\,odd} &=& 
\frac{\left(1-(-1)^m\right)}{8 f^2 \pi ^2} ({q_j}+{q_k})
  \sum_{j=0,\rm even}^{m}     a^{(3)}_{m+1,j}
\\&&\hspace*{5mm}
\times\Bigg\{ \frac{(u\cdot\Delta )^{m+1}}{24}
  \int_{-1}^{1}\frac{d\alpha}{\alpha^j} \Bigg[\frac{8 \left(2 \alpha ^j+3
        \left(\alpha 
   ^2-1\right) \alpha ^{m+1}\right) M_{\pi }^2 \delta
_{{uj}}^4}{\left(4 M_{\pi}^2+t \left(\alpha
    ^2-1\right)\right)^2}
-\frac{24 
   \left(\alpha ^j-\alpha ^{m+1}\right) M_{\pi }^2 \delta
   _{{uj}}^2}{4 M_{\pi}^2+t \left(\alpha ^2-1\right)}
\nonumber\\&&\hspace*{1.5cm}
+ \log 
   \left(\frac{M_{\pi }^2+\frac{1}{4} t \left(\alpha ^2-1\right)}{\mu
       ^2}\right) G(M_\pi)
- \log 
   \left(\frac{M_{{uj}}^2+\frac{1}{4} t \left(\alpha
         ^2-1\right)}{\mu ^2}\right) G(M_{uj})\Bigg]
\nonumber  
\\&&\nonumber\hspace*{10mm}
+
  \left[(u\cdot \Delta )^{m+1}-(2\overline{P}\cdot
     u)^{m+1-j} (u\cdot \Delta )^j \right]
 \left(\log
   \left(\frac{M_{\pi }^2}{\mu ^2}\right) M_{\pi }^2-\log
   \left(\frac{M_{{uj}}^2}{\mu ^2}\right) M_{{uj}}^2\right)
\Bigg\}\,,
\end{eqnarray}
where $G(M)= \left(t \left(3 \alpha ^2-7\right) \alpha ^j+t
  \left((j-m-4) \alpha ^2-j+m+8\right) \alpha ^{m+1}+4 \left(\alpha
    ^j+(j-m-2) \alpha ^{m+1}\right) M^2\right)$.  Note that this
matrix element vanishes in the QCD limit and the sea-isospin limit
where $q_j=-q_k$, in which foreseeable lattice calculations would be
performed. Consequently, we do not present the form-factors that
result.

\subsection{Tensor operators}
\label{sec:tens-oper}

For $m$ odd, the matrix elements of the tensor operator are given by:
\begin{eqnarray}
  \langle \pi^+|{\cal O}^{(m)}_{T}|\pi^+\rangle
  &=&   
   8  (1-(-1)^m)\overline{P}\cdot u r\cdot \Delta
  \sum_{j=0,\rm even}^{m}b_{m+1,j}(2\overline{P}\cdot u)^{m-j} (u\cdot \Delta )^j \Bigg\{
(1+y)\left[1+\frac{M_{uj}^2}{12\pi^2f^2}\log\left(\frac{M_{uj}^2}{\mu^2}\right)\right]
\nonumber\\&&\hspace*{1.5cm}
-\frac{1}{48 f^2 \pi ^2} 
 \Bigg[2 (3 ({q_3}+ {q_4})+5( 1+y)) \log
   \left(\frac{M_{{uj}}^2}{\mu ^2}\right) M_{{uj}}^2
\label{eq:22}
\\&&\hspace*{3cm}\nonumber
-3 \log
   \left(\frac{M_{\pi 
   }^2}{\mu ^2}\right) \left((2 ({q_3}+{q_4})-(1+y)) M_{\pi
 }^2+(1+y) \delta _{{uj}}^2\right)\Bigg]\Bigg\}\,,
\end{eqnarray}
which simplifies considerably in the QCD limit.  These matrix elements
vanish if the charge matrix is isovector in both the sea and valence
sectors. Matrix elements in the $\pi^-$ and $\pi^0$ are identical.

For $m$ even, the matrix elements are given by
\begin{eqnarray}
  \langle \pi^+|{\cal O}^{(m)}_{T}|\pi^+\rangle
  &=&   
  2 (1-y) \overline{P}\cdot u r\cdot\Delta \sum_{j=0,\rm
    even}^{m}  b_{m+1,j}\Bigg\{-4 (u\cdot \Delta )^j 
  \left((p\cdot u)^{m-j}-2 (2\overline{P}\cdot u)^{m-j}+((p+\Delta)\cdot
    u)^{m-j}\right) 
\nonumber
\\&&\hspace*{1cm}\times
\Bigg[1+\frac{M_{uj}^2}{12\pi^2f^2}\log\left(\frac{M_{uj}^2}{\mu^2}\right)
+\frac{1}{48 f^2 \pi ^2}
\left[3 \log \left(\frac{M_{\pi }^2}{\mu ^2}\right) (M_{\pi }^2+
  \delta _{{uj}}^2) -10 \log
   \left(\frac{M_{uj}^2}{\mu ^2}\right) M_{uj}^2\right]\Bigg]
\nonumber\\&&\nonumber
- \frac{(u\cdot \Delta )^m}{16 f^2 \pi ^2}
\int_{-1}^{1}d\alpha
 \left(4  M_{{uj}}^2+t \left(\alpha ^2-1\right)\right)
\\&&\hspace*{1cm}
\times\left[
   \left(\frac{1-\alpha}{2} \right)^{m-j}-2 \alpha ^{m-j}+
   \left(\frac{1+\alpha}{2}\right)^{m-j}\right] 
  \log \left(\frac{M_{{uj}}^2+\frac{1}{4} t
       \left(\alpha ^2-1\right)}{\mu ^2}\right)
\Bigg\}\,,
  \label{eq:23}
\end{eqnarray}
vanishing in the isoscalar case. The $\pi^-$ matrix elements differ by
an overall sign and the $\pi^0$ matrix elements vanish.

The results in Eqs.~(\ref{eq:22}) and (\ref{eq:23}) are easily
converted into the various GPD form factors. For $m$ odd,
\begin{eqnarray}
  \label{eq:24}
  B^{T}_{m+1,j}(t)&=&
B^{T,(0)}_{m+1,j}\Bigg\{\frac{(1+y)}{2}\left[
1- \frac{M_{uj}^2}{8\pi^2 f^2} \log
   \left(\frac{M_{{uj}}^2}{\mu ^2}\right)
  - \frac{M_\pi^2-\delta_{uj}^2}{16\pi^2 f^2}\log
   \left(\frac{M_{\pi 
   }^2}{\mu ^2}\right)
\right]
-\frac{q_3+q_4}{16\pi^2 f^2}M_{uj}^2 \log\left(\frac{M_{{uj}}^2}{\mu ^2}\right)
\nonumber \\&&\hspace*{15mm}
+\frac{q_3+q_4}{16\pi^2 f^2}M_{\pi}^2 \log\left(\frac{M_{{\pi}}^2}{\mu
    ^2}\right)
\Bigg\}
+B^{T,(a)}_{m+1,j}M_\pi^2
+\tilde{B}^{T,(a)}_{m+1,j}M_{uj}^2
+B^{T,(b)}_{m+1,j}t\,,
\end{eqnarray}
for all $j$. Here $B^{T,(0)}_{m+1,j}=8\,2^{m-j}\Lambda_\chi b_{m+1,j}$
($m$ odd).

While for $m$ even,
\begin{eqnarray}
  B^{T}_{m+1,j}(t)&=&
B^{T,(0)}_{m+1,j}\frac{1-y}{2}\Bigg\{
1+\frac{1}{16 \pi ^2 f^2 } 
 \Bigg[\left(M_\pi^2+\delta_{uj}^2\right)
\log\left(\frac{M_\pi^2}{\mu^2}\right)
-2M_{uj}^2 \log\left(\frac{M_{uj}^2}{\mu^2}\right) \Bigg]\Bigg\}
\nonumber
\\&&\hspace*{1cm}
+B^{T,(a)}_{m+1,j}M_\pi^2
+\tilde{B}^{T,(a)}_{m+1,j}M_{uj}^2
+B^{T,(b)}_{m+1,j}t\,,
  \label{eq:25}
\end{eqnarray}
for $j<m$, and 
\begin{eqnarray}
    B^{T}_{m+1,m}(t)&=&
B^{T,(0)}_{m+1,m}\frac{1-y}{2}\Bigg\{
1+\frac{1}{16 \pi ^2 f^2 } 
 \Bigg[\left(M_\pi^2+\delta_{uj}^2\right)
\log\left(\frac{M_\pi^2}{\mu^2}\right)
-2M_{uj}^2 \log\left(\frac{M_{uj}^2}{\mu^2}\right) \Bigg]\Bigg\}
\nonumber
\\&&
+\frac{1-y}{32 f^2 \pi ^2}
\sum_{j=0,\rm even}^{m}  B^{T,(0)}_{m+1,j}
\int_{-1}^{1}d\alpha
 \left(M_{{uj}}^2+\frac{t}{4} \left(\alpha ^2-1\right)\right)
\left[\frac{\alpha}{2} \right]^{m-j}
  \log \left(\frac{M_{{uj}}^2+\frac{1}{4} t
       \left(\alpha ^2-1\right)}{\mu ^2}\right)
\nonumber\\
&&+B^{T,(a)}_{m+1,m}M_\pi^2
+\tilde{B}^{T,(a)}_{m+1,m}M_{uj}^2
+B^{T,(b)}_{m+1,m}t\,,
  \label{eq:26}
\end{eqnarray}
where $B_{m+1,j}^{T,(0)} = -8\Lambda_\chi\left[
  \sum_{k=0,\rm even}^{m}\left( {\tiny \begin{array}{c} m-k \\
        j-k\end{array}}\right) 2^{k-j}
  b_{m+1,k}-2^{m-j}b_{m+1,j}\right]$ ($m$ even) is the leading order
result for the form-factor. In the QCD limit, these result reproduce
those of Ref.~\cite{Diehl:2006js} [Eqs.~(106) and (128) therein]. As
these results are insensitive to the sea-quark charges, these
form-factors can be calculated in lattice QCD without disconnected
quark loops.

\section{Finite Volume Effects}
\label{sec:finite-volume-funct}

The space-time lattices used in numerical simulations are of finite
extent by necessity, and consequently, lattice results differ from
those of the physical (infinite volume) world even at the physical
quark masses.\footnote{Lattice artifacts from the discretisation of
  space-time also influence lattice results. We do not discuss these
  here and assume a continuum extrapolation has been performed.} For
sufficiently large volumes, these effects can be incorporated into the
effective field theory approach, allowing infinite volume results to
be extracted from the finite volume (FV) lattice simulations.  Here we
shall consider a hyper-rectangular box of dimensions $L^3\times T$
with $T\gg L$ as appropriate for most current lattice calculations.
Imposing periodic boundary conditions on mesonic fields leads to
quantised momenta $k=(k_0,{\vec k})$, ${\vec k}=\frac{2\pi}{L} {\vec
  j}=\frac{2\pi}{L} (j_1,j_2,j_3)$ with $j_i\in \mathbb{Z}$, but $k_0$
treated as continuous.  On such a finite volume, spatial momentum
integrals are replaced by sums over the available momentum modes. This
leads to modifications of the infinite volume results presented in the
previous section; the various functions arising from loop integrals
are replaced by their FV counterparts. In a system where $M_\pi L\gg
1$, finite volume effects are predominantly from Goldstone mesons
propagating to large distances where they are sensitive to boundary
conditions and can even ``wrap around the world''.\footnote{In
  principle, finite volume effects can also be computed for $M_\pi
  L\sim1$ but $\Lambda_\chi L\gg1$ where mesonic zero-modes become
  enhanced \cite{Leutwyler:1987ak,Gasser:1987zq,Detmold:2004ap}.
  These calculations are beyond the scope of this work.}  Since the
lowest momentum mode of the Goldstone propagator is $\sim \exp(-M_\pi
L)$ in position space, finite volume effects will behave as a
polynomial in $1/L$ times this exponential provided no multi-particle
thresholds are reached, that is for $t < 4 M_\pi^2$. For $t\geq
4M_\pi^2$, volume effects that are polynomial in inverse powers of $L$
are expected, however for realistic lattice calculations, such
momentum transfers are too large to be described in \xpt\ and we
neglect the resulting complications in our analysis.

The finite volume effects in the diagrams of
Fig.~\ref{fig:XPT-diagrams}(b) and (c) arising from the operators that
contribute at infinite volume are well known. However the effects in
Fig.~\ref{fig:XPT-diagrams}(a) are made more complicated by the
non-trivial numerator structure of the integral/sum and the requisite
details for their evaluation are given in Appendix \ref{sec:FVapp}.
The final forms for the finite volume versions of our results are
given below.
In most cases, these replacements of spatial momentum integrals by
sums would complete the calculation of finite volume effects. However
for the GPD form-factors, there are additional complications that
arise from operators whose contributions vanish in infinite volume but
which give non-zero contributions at finite volume. To see this, we
concentrate on the vector case and reconsider the operator in
Eq.~(\ref{eq:14}) (note there is some redundancy of terms in this
operator).  Na\"\i{}vely, this operator would contribute through
diagram (b) in Fig.~\ref{fig:XPT-diagrams}.  As discussed in
Section~\ref{sec:effect-field-theory}, there is no contribution at
infinite volume as the each of the pion fields coming from this
operator must have a non-zero number of derivatives acting on it.
Combining $u^2=0$ (or equivalently, the definite Lorentz
transformation properties of the twist-two, spin-$n$, operators) and
the symmetric range of integration, terms with derivatives inside the
loop in Fig.~\ref{fig:XPT-diagrams}(b) give no contribution.

In the finite volumes discussed here, Lorentz symmetry (or O(4)
symmetry in Euclidean space as relevant in lattice calculations) is
explicitly broken by the imposition of the boundary conditions (in a
$L^3\times T$ periodic box, arbitrary Lorentz boosts do not leave the
system invariant).  As a result, the arguments used to discard the
operators in Eq.~(\ref{eq:14}) break down ($\frac{1}{L^3}\int
dq_0\sum_{\vec{q}} \frac{(u.q)^2}{(q^2-m^2)^j}\ne 0$ at finite volume)
and their contributions, which must vanish as $L\to\infty$, need to be
incorporated to correctly describe finite volume lattice calculations
Consequently, one must include the functional dependence arising from
operators such as those in Eq.~(\ref{eq:14}) in any fit to lattice
data and determine the relevant combinations of the
$a^{A,2}_{m,j_1,j_2,j_3}$ as well as the $a_{m+1,j}^A$ before
discarding the former (we have no interest in these LECs for the
matrix elements we want to extract, however they would contribute in
more complicated matrix elements such as two pion matrix elements of
twist-two operators) in extracting the infinite volume result.  In
general this is a very intricate task as the functional dependence
produced by these operators in the contribution of
Fig.~\ref{fig:XPT-diagrams}(b) to the matrix element of ${\cal
  O}^{(m)}_A$ is
\begin{eqnarray}
  \label{eq:27}
  \sum_{l=2,\,{\rm even}}^{n} \gamma_{n,l}u_{\mu_1}\ldots u_{\mu_l}\int
  d k_0 \frac{1}{L^3} 
  \sum_{\vec{k}} \frac{k^{\mu_1}\ldots k^{\mu_l}-{\rm tr}}{k^2-m^2}
\end{eqnarray}
where the parameters $\gamma_{n,l}$ are linear combinations of those
that enter in Eq.~(\ref{eq:14}).\footnote{Care must to be taken to
  define a suitable regularisation prescription.}  However for the low
moments ($m<4$) that are accessible in current lattice calculations,
these subtractions can be performed.

A number of additional aspects of Lorentz symmetry violation at finite
volume are worth noting. Firstly, these effects did not contribute in
the nucleon twist-two matrix elements at finite volume at NLO
\cite{Detmold:2005pt}, but will contribute at higher orders in the
chiral expansion in a similar way as they enter here. We also note
that moments of distribution amplitudes (given by meson to vacuum
matrix elements of related quark-bilinear operators) will suffer from
similar complications at finite volume and the absence of non-analytic
quark mass dependence at NLO found in Ref.~\cite{Chen:2003fp} at
infinite volume will not persist. Secondly, as shown in
Ref.~\cite{Gasser:1987zq}, no new operators can appear at finite
volume otherwise their LECs would necessarily depend on $L^{-1}$, an
infrared scale. In the case described above, the operator is present
both at finite volume and infinite volume, but doesn't contribute to
the matrix elements we consider in the latter case. A final related
effect of the finite volume is that Lorentz symmetry can no longer be
used to decompose the matrix elements of the twist-two operators. At
finite volume, additional structures that violate Lorentz symmetry can
appear in the decomposition of the matrix elements in
Eqs.~(\ref{eq:7})--(\ref{eq:10})\footnote{In the EFT analysis of the
  matrix elements under consideration, such terms do not appear until
  higher orders (two-loops) in the chiral expansion, but even to the
  order we work, form-factors acquire dependence on $|\vec{u}|^2$.}.
The form factors of such terms vanish as $L\to\infty$ and we ignore
them here but care must be taken to correctly extract the form-factors
in Eqs.~(\ref{eq:7}--\ref{eq:10}) without pollution from these
additional Lorentz non-invariant form-factors.

\subsection{Finite volume matrix elements}

With the preceding remarks in mind, the finite volume versions of the
results of the previous section are given by the following matrix
elements:
\begin{eqnarray}
  \label{eq:28}
  \langle \pi^+|{\cal O}^{(m)}_{0}|\pi^+\rangle^{m\,\rm odd}_{\rm FV}
  &=&   
2(1-(-1)^m) \sum_{j=0, {\rm even}}^{m}  a^{(0)}_{m+1,j} \Bigg\{\left(
  (2\overline{P}\cdot u)^{m+1-j} (u\cdot 
   \Delta )^j-(u\cdot \Delta )^{m+1}\right)
\\&&
+\frac{ (u\cdot \Delta )^{m+1}}{3 f^2}
\Bigg[-2
   \left(2M_{\pi }^2-\delta_{uj}^2+2 t\right)
   K_{0,0,0,0,0}^{{uj}}
\nonumber\\&&\hspace*{5mm}
+3 M_{\pi }^2 
   \left(K_{0,0,0,0,0}^{{uu}}+\delta _{{uj}}^2
   \left(K_{0,0,0,0,0}^{{uu1}}+K_{0,0,0,0,0}^{{uu2}}\right)\right)+4
   \left(K_{0,0,0,1,0}^{{uj}}+K_{0,1,0,0,0}^{{uj}}\right) 
\nonumber\\&&\hspace*{5mm}
- 
   \sum_{k=0}^{m+1-j} \left( {\tiny \begin{array}{c} m+1-j \\
           k\end{array}}\right) 2^k (u\cdot \Delta )^{-k} 
\Big(-2 \left(2M_{\pi }^2-\delta_{uj}^2+2 t\right)
   K_{k,0,0,0,0}^{{uj}}
\nonumber\\&&\hspace*{5mm}
+3 M_{\pi }^2
   \left(K_{k,0,0,0,0}^{{uu}}+\delta 
   _{{uj}}^2
   \left(K_{k,0,0,0,0}^{{uu1}}+K_{k,0,0,0,0}^{{uu2}}\right)\right)+4
   \left(K_{k,0,0,1,0}^{{uj}}
     +K_{k,1,0,0,0}^{{uj}}\right)\Big)\Bigg]\Bigg\}\,,
\nonumber
\end{eqnarray}
\begin{eqnarray}
  \label{eq:29}
  \langle \pi^+|{\cal O}^{(m)}_{3}|\pi^+\rangle^{m\,\rm even}_{\rm FV}
  &=&  
2(1+(-1)^m) \sum_{j=0,{\rm even}}^{m} a^{(3)}_{m+1,j}\Bigg\{ (2\overline{P}\cdot u)^{m+1-j} (u\cdot \Delta )^j
   \left(1-\frac{2}{f^2} J_{u,j}\right)
\\&&\nonumber\hspace*{40mm}
-\frac{2 }{f^2}
   \sum _{k=0}^{m+1-j}
\left( {\tiny \begin{array}{c} m+1-j \\ k\end{array}}\right) 
 2^k  (u\cdot \Delta )^{m+1-k} \left(K_{k,0,0,1,0}^{{uj}}+2
   K_{k,0,1,0,0}^{{uj}}\right)
\Bigg\}\,,
\end{eqnarray}
\begin{eqnarray}
  \label{eq:30}
  \langle \pi^+|{\cal O}^{(m)}_{3}|\pi^+\rangle^{m\,\rm odd}_{\rm FV}
  &=& 
\frac{\left(1-(-1)^m\right)}{3 f^2}\left(q_j+q_k\right)\sum_{j=0,{\rm even}}^{m} a^{(3)}_{m+1,j}
   \Bigg\{(u\cdot \Delta )^{m+1} 
\Big[-\left(2 M_{\pi }^2-\delta _{{uj}}^2+2
   t\right) K_{0,0,0,0,0}^{{uj}}
\\&&\hspace*{10mm}
+2 \left(M_{\pi }^2+t\right)
   K_{0,0,0,0,0}^{{uu}}
+3 M_{\pi }^2 \delta _{{uj}}^2
   \left(K_{0,0,0,0,0}^{{uu1}}+K_{0,0,0,0,0}^{{uu2}}+\delta _{{uj}}^2
   K_{0,0,0,0,0}^{{uu3}}\right)
\nonumber\\&&\hspace*{10mm}
+2
   \left(K_{0,0,0,1,0}^{{uj}} -K_{0,0,0,1,0}^{{uu}} 
+K_{0,1,0,0,0}^{{uj}}-K_{0,1,0,0,0}^{{uu}}\right)\Big]
\nonumber\\&&
-\sum _{k=0}^{m+1-j}
\left( {\tiny \begin{array}{c} m+1-j \\ k\end{array}}\right) 
2^k (u\cdot \Delta )^{m+1-k} \Big[-\left(2
   M_{\pi }^2-\delta _{{uj}}^2
+2 t\right) K_{k,0,0,0,0}^{{uj}}+2 \left(M_{\pi
   }^2+t\right) K_{k,0,0,0,0}^{{uu}}
\nonumber\\&&\hspace*{10mm}
+3 M_{\pi }^2 \delta _{{uj}}^2
   \left(K_{k,0,0,0,0}^{{uu1}} +K_{k,0,0,0,0}^{{uu2}} 
+\delta _{{uj}}^2
   K_{k,0,0,0,0}^{{uu3}}\right)
\nonumber\\&&\hspace*{10mm}
+2
   \left(K_{k,0,0,1,0}^{{uj}} -K_{k,0,0,1,0}^{{uu}}
     +K_{k,1,0,0,0}^{{uj}}-K_{k,1,0,0,0}^{{uu}}\right)\Big]
\nonumber\\&&
+6
   \left(2^{-j+m+1} (\overline{P}\cdot u)^{m+1-j} (u\cdot \Delta )^j-(u\cdot
     \Delta )^{m+1}\right) 
   \left(J_{u,j}-J_{u,u}\right)\Bigg\}
\nonumber    \,,
\end{eqnarray}
\begin{eqnarray}
  \langle \pi^+|{\cal O}^{(m)}_{T}|\pi^+\rangle^{m\,\rm even}_{\rm FV}
  &=& 
2 (1-y) \overline{P}\cdot u 
   r\cdot \Delta \sum_{j=0,{\rm even}}^{m} \frac{B^{T\,(0)}_{m+1,j}}{\Lambda_\chi}
 (u\cdot \Delta )^j
     (\overline{P}\cdot u)^{m-j}
   \left(1+\frac{1}{f^2}(L_{u,u} \delta _{{uj}}^2-2
       J_{u,j}+J_{u,u})\right) 
\nonumber\\&&
-
\frac{16 (1-y)}{f^2}\sum_{j=0,{\rm even}}^{m}
 b_{m+1,j}
\Bigg\{
- r\cdot \Delta  (u\cdot 
   \Delta )^{j} 
   \left(K_{m+1-j,0,0,1,0}^{{uj}}+2
     K_{m+1-j,0,1,0,0}^{{uj}}\right)
\nonumber\\&& \hspace*{2cm}
+(u\cdot \Delta )^{j+1}\left(K_{m-j,0,0,1,1}^{{uj}}+2
     K_{m-j,0,1,0,1}^{{uj}}\right)   
\nonumber\\&& \hspace*{1cm}
+\sum _{k=0}^{m-j}
\left( {\tiny \begin{array}{c} m-j \\ k\end{array}}\right)
\left(2^{k+1}-1\right) (u\cdot \Delta 
   )^{m-k} \Bigg[ r\cdot \Delta
   \left(K_{k+1,0,0,1,0}^{{uj}}+2
     K_{k+1,0,1,0,0}^{{uj}}\right)
\nonumber\\&& \hspace*{2cm}
-  u\cdot \Delta
   \left(K_{k,0,0,1,1}^{{uj}}+2 
   K_{k,0,1,0,1}^{{uj}}\right)
 \Bigg]\Bigg\}\,,
  \label{eq:31}
\end{eqnarray}
and
\begin{eqnarray}
  \label{eq:32}
  \langle \pi^+|{\cal O}^{(m)}_{T}|\pi^+\rangle^{m\,\rm odd}_{\rm FV} &=&      
8 r\cdot \Delta \sum_{j=0,{\rm even}}^{m}
 (2\overline{P}\cdot u)^{m+1-j} (u\cdot \Delta )^j
   b_{m+1,j} \Bigg[(y+1) \left(1-\frac{1}{f^2}(2 J_{u,j}+J_{u,u}-\delta _{{uj}}^2
   L_{u,u})\right)
\\&&\hspace*{2cm}
-\frac{2}{f^2} \left(q_3+q_4\right) \left(J_{u,j}-
   J_{u,u}\right)\Bigg]
\nonumber\\&&
-\frac{16 (u\cdot \Delta )^{m}}{3 f^2} \sum_{j=0,{\rm
    even}}^{m}  b_{m+1,j}
\sum _{k=0}^{m-j}
\left( {\tiny \begin{array}{c} m-j \\ k\end{array}}\right) 2^k(u\cdot \Delta )^{-k}
\nonumber\\&&\hspace*{2mm}
\times\Bigg\{
u\cdot \Delta  \Bigg[
\left(2M_\pi^2-\delta_{uj}^2+2
   t\right) {\cal Q}_{34} K_{k,0,0,0,1}^{{uj}}
-2{\cal Q}_{34} K_{k,0,0,1,1}^{{uj}}
-2{\cal Q}_{34} K_{k,1,0,0,1}^{{uj}}
\nonumber\\&&\hspace*{1cm}
+M_{\pi }^2
   \Big({\cal Q}_{34} K_{k,0,0,0,1}^{{uj}}
- {\cal Q}_{34}^{\prime\prime}K_{k,0,0,0,1}^{{uu}}
-3 \left(q_3+q_4\right) \delta
   _{{uj}}^2
   \left(K_{k,0,0,0,1}^{{uu1}}+K_{k,0,0,0,1}^{{uu2}}\right)
\nonumber\\&&\hspace*{1cm}
+3 {\cal Q}_{34}^\prime \delta_{{uj}}^4 K_{k,0,0,0,1}^{{uu3}}\Big)
-2{\cal Q}_{34}^\prime
 \left(-t K_{k,0,0,0,1}^{{uu}} +K_{k,0,0,1,1}^{{uu}}
   +K_{k,1,0,0,1}^{{uu}}\right)
\Bigg]
\nonumber\\&&\hspace*{5mm}
-r\cdot \Delta  \Bigg[\left(2 M_{\pi}^2-\delta_{uj}^2+2 t\right) 
   {\cal Q}_{34} K_{k+1,0,0,0,0}^{{uj}}
-2{\cal Q}_{34} K_{k+1,0,0,1,0}^{{uj}}
  -2{\cal Q}_{34} K_{k+1,1,0,0,0}^{{uj}}
\nonumber\\&&\hspace*{1cm}
+M_{\pi }^2
   \Big({\cal Q}_{34} K_{k+1,0,0,0,0}^{{uj}}
     -{\cal Q}_{34}^{\prime\prime} K_{k+1,0,0,0,0}^{{uu}}
-3 \left(q_3+q_4\right) \delta_{{uj}}^2
  \left(K_{k+1,0,0,0,0}^{{uu1}}
    +K_{k+1,0,0,0,0}^{{uu2}}\right)
\nonumber\\&&\hspace*{1cm}
     +3 {\cal Q}_{34}^\prime \delta_{{uj}}^4 K_{k+1,0,0,0,0}^{{uu3}}\Big)
  -2{\cal Q}_{34}^\prime \left(-t K_{k+1,0,0,0,0}^{{uu}}
 +K_{k+1,0,0,1,0}^{{uu}}
   +K_{k+1,1,0,0,0}^{{uu}}\right)
\Bigg]
\Bigg\}\,,
\nonumber
\end{eqnarray}
where ${\cal Q}_{34}=1+y+q_3+q_4$, ${\cal Q}_{34}^\prime=1+y-q_3-q_4$
and ${\cal Q}_{34}^{\prime\prime}=1+ y+2 q_3+2 q_4$. Note that the
entire last sum in the above expression vanishes identically at
infinite volume.

Here 
\begin{eqnarray}
  \label{eq:33}
  J_{a,b}=\int\frac{dq_0}{2\pi}\frac{1}{L^3}\sum_{\vec{q}}
\frac{i}{q^2-M_{ab}^2+i\epsilon},
&\quad\quad&
  L_{a,b}=\int\frac{dq_0}{2\pi}\frac{1}{L^3}\sum_{\vec{q}}
\frac{i}{(q^2-M_{ab}^2+i\epsilon)^2},
\end{eqnarray}
and
\begin{eqnarray}
  \label{eq:34}
  K_{i,a,b,c,d}^{fg} ={\cal K}_{i,a,b,c,d}^{f,g;1,1}, \quad
  K_{i,a,b,c,d}^{fg1} ={\cal K}_{i,a,b,c,d}^{f,g;2,1}, \quad
  K_{i,a,b,c,d}^{fg2} ={\cal K}_{i,a,b,c,d}^{f,g;1,2}, \quad
  K_{i,a,b,c,d}^{fg3} ={\cal K}_{i,a,b,c,d}^{f,g;2,2}, \quad
\end{eqnarray}
where
\begin{eqnarray}
  \label{eq:35}
 {\cal K}_{i,a,b,c,d}^{f,g;n_1,n_2} &=&
  -i\int\frac{dq_0}{2\pi}\frac{1}{L^3}\sum_{\vec{q}}
\frac{(u\cdot q)^i (q^2)^a (p\cdot q)^b (q\cdot\Delta)^c (q\cdot r)^d}
{(q^2-M_{fg}^2+i\epsilon)^{n_1}((q+\Delta)^2-M_{fg}^2+i\epsilon)^{n_2}}
\end{eqnarray}
which in turn is simply represented in terms of the functions
appearing in Appendix \ref{sec:FVapp}.

Since the external momenta $p=\overline{P}-\Delta/2$, $\Delta$, $u$
and $r$ occur in the various integrals/sums, it is not possible to
decompose these matrix elements into form-factors without specifying
to a particular $m$.

\section{Discussion}
\label{sec:discussion}

The results for the quark mass and volume dependence of the moments of
the pion GPDs presented above are useful in extrapolating lattice data
at small $t$ in the chiral regime (the range of volumes and pion
masses for which the chiral expansion is convergent) to the physical
world. Although subtleties not seen in simpler quantities arise at
finite volume that complicate the extraction of the infinite volume
results for the twist-two matrix elements, we have outlined a
procedure by which the extrapolation can be performed reliably. The
knowledge of the meson GPDs that can be obtained by combining our
results with lattice calculations will be useful in comparison to, and
as a complement to, experiment.

\acknowledgements

We are grateful to C.-J.~D.~Lin for discussions. J.-W.~C. is supported
by the National Science Council of R.O.C. and W.~D. and B.~S. are
supported by US Department of Energy grant DE-FG02-97ER41014. W.~D. is
grateful to the Department of Physics at National Taiwan University
and the National Center for Theoretical Sciences, Taiwan for
hospitality.

\appendix

\section{Finite volume functions}
\label{sec:FVapp}

The functions used in our results at finite volume can be built from
the following basic structure
\begin{eqnarray}
  \label{eq:36}
{\cal J}_{\alpha,n_1,n_2}^{i_1\ldots i_\beta}(m_1,m_2,\Delta;L)&=&
-i  \int\frac{d\,q_0}{2\pi}\frac{1}{L^3}\sum_{\vec{q}}
\frac{q_0^{2\alpha}\vec{q}^{i_1}\ldots \vec{q}^{i_\beta}}
{(q^2-m_1^2+i\epsilon)^{n_1}
  ((q+\Delta)^2-m_2^2+i\epsilon)^{n_2}}   
\nonumber \\
&=&
 -\frac{1 }{L^3\,\Gamma(n_1)\Gamma(n_2)}
\int_0^1 d\,x    \sum_{\vec{k}}
\frac{\partial^{n_1+n_2-1}}{\partial k_0^{n_1+n_2-1}}
\left[\frac{k_0^{2\alpha}\vec{k}^{i_1}\ldots \vec{k}^{i_\beta}}
{\left[k_0 +\hat{k}_0\right]^{n_1+n_2}}\right]_{k_0=\hat{k}_0}\,,
\end{eqnarray}
where 
\begin{displaymath}
  \hat{k}_0=\sqrt{|\vec{k}|^2 + 2(1-x)\vec{k}\cdot\vec\Delta +
  (1-x)|\vec\Delta|^2+ x m_1^2 +(1-x) m_2^2}\,,
\end{displaymath}
after performing the energy integral by contour integration. 
As an example, we find that
\begin{eqnarray}
  \label{eq:37}
  K^{ab}_{j,0,0,0,1}&=&
  \sum_{k=0}^{\lfloor \frac{j}{2}\rfloor}
 (-1)^{j-k-1}\left[ u_0^{2k+1}\vec{u}^{i_1}\ldots\vec{u}^{i_{j-k-1}}
r_0 {\cal J}_{k+1,1,1}^{i_1,\ldots,i_{j-k-1}}
-u_0^{2k}\vec{u}^{i_1}\ldots\vec{u}^{i_{j-k}}
\vec{r}^{i_{j-k+1}} {\cal J}_{k,1,1}^{i_1,\ldots,i_{j-k+1}}\right]\,.
\end{eqnarray}
It is possible to write down a general expression for the multiple
derivative in Eq.~(\ref{eq:36})\footnote{We may use the well known
  formula of Fa\`a di Bruno
\begin{equation}
  \label{eq:38}
  \frac{d^n}{d\,x^n}g\left(f(x)\right)
  =\sum_{k_1=0}^{n}\ldots\sum_{k_n=0}^{n}  
  \frac{n!\delta(n-\sum_{i=1}^n i\, k_i)}{\prod_{i=1}^n k_i!}
    \frac{d^{K}g(f(t))}{d\,f(t)^K}\prod_{i=1}^n
    \left(\frac{f^{(i)}(t)}{i!} \right)^{k_i}  \,,
\end{equation}
where $K=\sum_{i=1}^n k_i$.} but since we only require
$\{n_1,n_2\}=\{1,1\},\{1,2\},\{2,1\},\{2,2\}$, it is easier to present
each case explicitly. This leads to
\begin{eqnarray}
  \label{eq:39}
  {\cal J}_{\alpha,1,1}^{i_1\ldots i_\beta}(m_1,m_2,\Delta;L)
&=& -\frac{ (2\alpha-1)}{4L^3\,}
\int_0^1 d\,x  \, I_{\frac{3}{2}-\alpha}^{i_1\ldots
  i_\beta}\left[(1-x)\vec\Delta, {\cal M}\right]\,,
\\  {\cal J}_{\alpha,1,2}^{i_1\ldots i_\beta}(m_1,m_2,\Delta;L)&=&
 -\frac{ (4\alpha^2-8\alpha+3)}{8L^3\,}
\int_0^1 d\,x \,(1-x)  I_{\frac{5}{2}-\alpha}^{i_1\ldots
  i_\beta}\left[(1-x)\vec\Delta, {\cal M}\right]\,,
\\
  {\cal J}_{\alpha,2,1}^{i_1\ldots i_\beta}(m_1,m_2,\Delta;L)&=&
 -\frac{ (4\alpha^2-8\alpha+3)}{8L^3\,}
\int_0^1 d\,x \,x\,  I_{\frac{5}{2}-\alpha}^{i_1\ldots
  i_\beta}\left[(1-x)\vec\Delta, {\cal M}\right]\,,
\\
  {\cal J}_{\alpha,2,2}^{i_1\ldots i_\beta}(m_1,m_2,\Delta;L)&=&
 -\frac{ (8 \alpha ^3-36 \alpha ^2+46 \alpha -15)}
{16L^3\,}
\int_0^1 d\,x \, x(1-x) I_{\frac{7}{2}-\alpha}^{i_1\ldots
  i_\beta}\left[(1-x)\vec\Delta, {\cal M}\right]\,,
\end{eqnarray}
where ${\cal M}=x(1-x)|\vec\Delta|^2 + x m_1^2 +(1-x) m_2^2$ and 
\begin{eqnarray}
  \label{eq:40}
  I_\beta^{i_1\ldots i_n}(\vec{z},m) &\equiv&
 \sum_{\vec{q}}\frac{q^{i_1}\ldots
    q^{i_n}}{\left(|\vec{q}+\vec{z}|^2+m^2\right)^\beta}\,.
\end{eqnarray}
These sums can then be further simplified using the recurrence
identity
\begin{eqnarray}
  \label{eq:41}
  I_\beta^{i_1\ldots i_n}(\vec{z},m)
  &=& -\frac{1}{2(\beta-1)}\frac{d I_{\beta-1}^{i_1\ldots i_{n-1}}}{d
    z_{i_n}} -z^{i_n} I_\beta^{i_1\ldots i_{n-1}}(\vec{z},m)
\nonumber \\
  &\longrightarrow &
(-1)^n \sum_{k=0}^{n}\sum_{\cal P} {\cal
  P}\left(\partial_z^{i_1}\ldots\partial_z^{i_k} z^{i_{k+1}}\ldots
  z^{i_n}\right) I_{\beta-k}(\vec{z},m)\,,
\end{eqnarray}
where ${\cal P}$ denotes a permutation of the orderings of the partial
derivatives, $\partial_z^{i}=\frac{\partial}{\partial z^i}$, and the
$z^i$'s, and the sum is over all such permutations. The remaining
momentum sums have scalar summands and are simple to evaluate using
the results of Ref.~\cite{Sachrajda:2004mi} and the tracelessness
condition. Divergences can be regulated using Epstein-Hurwitz
zeta-function techniques \cite{Elizalde:1997jv}.

\section{QCD form-factors}
\label{sec:qcd-form-factors}

For easy reference, the non-vanishing, infinite volume form-factors in
the SU(2) theory are given here.

In the vector-isoscalar case for $m$ odd,
\begin{eqnarray}
  \label{eq:42}
 A^{(0)}_{m+1,j}(t)&=&A^{(0)}_{m+1,j}+(A^{(0,a)}_{m+1,j}
+\tilde{A}^{(0,a)}_{m+1,j})M_{\pi}^2
+A^{(0,b)}_{m+1,j}t, \ \text{for} \ j \leq m \,,  
\end{eqnarray}
and
\begin{eqnarray}
  \label{eq:43}
  C^{(0)}_{m+1}(t)&=&C^{(0)}_{m+1}
+\frac{C^{(0)}_{m+1}}{96\pi^2f^2}\int_{-1}^{1} 
d\alpha 
((3\alpha^2-7)t+7M_{\pi}^2)
\log\left[\frac{M_{\pi}^2-\frac{t}{4}(1-\alpha^2)}{\mu^2}
\right]
+(C^{(0,a)}_{m+1}+\tilde{C}^{(0,a)}_{m+1})M_\pi^2+C^{(0,b)}_{m+1}t 
\nonumber  \\ &&
+\frac{1}{192 \pi^2 f^2}\sum_{j=0,\text{even}}^{m}
2^{j-m} A^{(0)}_{m+1,j}
\\&&\hspace*{0.6cm}
\times \int_{-1}^{1} d\alpha \
\alpha^{m-j+1}\left(t(3\alpha^2-7)+7M_{\pi}^2+4(m-j+1)
\left(M_{\pi}^2-\frac{t}{4}(1-\alpha^2)\right)\right)
\log\left[\frac{M_{\pi}^2-\frac{t}{4}(1-\alpha^2)}{\mu^2}\right]
\,.
\nonumber
\end{eqnarray}

In the vector-isovector case for $m$ even,
\begin{eqnarray}
  \label{eq:44}
  A^{(3)}_{m+1,j}(t) &=& A^{(3)}_{m+1,j}
  \left(1-\frac{M_{\pi}^2}{8\pi^2f^2}
    \text{log}\left(\frac{M_{\pi}^2}{\mu^2}\right)
  \right)+(A^{(3,a)}_{m+1,j}+\tilde{A}^{(3,a)}_{m+1,j})M_\pi^2 
+A^{(3,b)}_{m+1,j}t, \ \text{for} \ j 
  \leq m-2  \,,  
\end{eqnarray}
and
\begin{eqnarray}
  \label{eq:45}
 A^{(3)}_{m+1,m}(t) &=& A^{(3)}_{m+1,m}
 \left(1-\frac{M_{\pi}^2}{8\pi^2f^2}
   \text{log}\left(\frac{M_{\pi}^2}{\mu^2}\right)\right) 
+(A^{(3,a)}_{m+1,m}+\tilde{A}^{(3,a)}_{m+1,m})M_\pi^2
+A^{(3,b)}_{m+1,m}t
 \\&& \nonumber
 +\sum_{j=0,\text{even}}^{m} \frac{2^{j-m}A_{m+1,j}^{(3)}}{8\pi^2 f^2}
\left\{M_{\pi}^2 \text{log} 
\left[\frac{M_{\pi}^2}{\mu^2}\right]
  -\frac{t}{4}
  \int_{-1}^{1}
 d\alpha \ \alpha^{m-j+2} \
 \text{log}\left[\frac{M_{\pi}^2-\frac{t}{4}(1-\alpha^2)}{\mu^2}
 \right]\right\}\,.
\end{eqnarray}

Finally in the isoscalar tensor case for $m$ odd
\begin{eqnarray}
  \label{eq:46}
  B^{T}_{m+1,j}(t)&=&
B^{T,(0)}_{m+1,j}
\left[
1- \frac{3M_{\pi}^2}{16\pi^2 f^2} \log
   \left(\frac{M_{{\pi}}^2}{\mu ^2}\right)
\right]
+(B^{T,(a)}_{m+1,j}
+\tilde{B}^{T,(a)}_{m+1,j})M_{\pi}^2
+B^{T,(b)}_{m+1,j}t\,,
\end{eqnarray}
while in the isovector tensor case for $m$ even,
\begin{eqnarray}
  \label{eq:47}
  B^{T}_{m+1,j}(t)&=&
B^{T,(0)}_{m+1,j}\left[
1-\frac{ M_\pi^2}{16 \pi ^2 f^2 } 
\log\left(\frac{M_\pi^2}{\mu^2}\right)
\right]
+(B^{T,(a)}_{m+1,j}
+\tilde{B}^{T,(a)}_{m+1,j})M_{\pi}^2
+B^{T,(b)}_{m+1,j}t\,,
\end{eqnarray}
for $j<m$, and 
\begin{eqnarray}
    B^{T}_{m+1,m}(t)&=&
B^{T,(0)}_{m+1,m}\left[
1-\frac{M_\pi^2}{16 \pi ^2 f^2 } 
\log\left(\frac{M_\pi^2}{\mu^2}\right)
\right]
+(B^{T,(a)}_{m+1,m}
+\tilde{B}^{T,(a)}_{m+1,m})M_{\pi}^2
+B^{T,(b)}_{m+1,m}t
\nonumber
\\&&
+\frac{1}{16 f^2 \pi ^2}
\sum_{j=0,\rm even}^{m}  B^{T,(0)}_{m+1,j}
\int_{-1}^{1}d\alpha
 \left(M_{{\pi}}^2+\frac{t}{4} \left(\alpha ^2-1\right)\right)
\left[\frac{\alpha}{2} \right]^{m-j}
  \log \left(\frac{M_{{\pi}}^2+\frac{1}{4} t
       \left(\alpha ^2-1\right)}{\mu ^2}\right)
\,.
  \label{eq:48}
\end{eqnarray}

\bibliography{pion_gpd}

\end{document}